\input{aipcheck}

\documentclass[
    ,final            
  ]
  {aipproc}

\layoutstyle{8x11double}

\begin{document}
\title{Possibility of a Two-Component Tomonaga-Luttinger Liquid 
in Frustrated Integer-Spin Tubes}



\author{Masahiro Sato}{
address={Synchrotron Radiation Research Center, 
Japan Atomic Energy Research Institute, Mikazuki, Sayo, Hyogo 679-5148, Japan and
CREST JST}
}



\begin{abstract}
Uniform-field effects 
for frustrated odd-leg integer-spin tubes (cylinder type spin 
systems) are investigated in the weak interchain-coupling regime. 
We predict that, as the field exceeds the spin gap, 
a two-component Tomonaga-Luttinger liquid (TLL) appears 
due to the condensation of the doubly degenerate lowest magnons. 
Furthermore, it is argued that when the uniform field is so strong that 
the second lowest magnons are also condensed, the two-component TLL 
is destroyed and a new one-component TLL emerges. 
This quantum phase transition may be detected as a magnetization cusp. 
\end{abstract}

\maketitle


In one-dimensional (1D) gapped isotropic spin systems, 
a uniform magnetic field, 
which causes the rotation symmetry to reduce to a $U(1)$ one, 
generally removes the gap and induces a magnon-condensed phase regarded as a
Tomonaga-Luttinger liquid (TLL)~\cite{Gogo} with central charge $c=1$.  
Zamolodchikov's $c$ theorem~\cite{Zamol} actually predicts that 
a $c=1$ TLL state tends to occur 
in $U(1)$ symmetric 1D systems. However, frustration often breaks such
conventional scenarios. 
In this short paper, we focus on the field-induced
critical phases in frustrated integer-spin tubes, and discuss the
possibility of the existence of an unconventional two-component TLL 
(we call this $c=1+1$ state in this paper) and a quantum phase transition. 
Here, the tube means a ladder with a periodic boundary condition 
along the interchain (rung) direction. For more details, refer to 
Ref.~\cite{M_S}.

The $N$(finite)-leg spin tube Hamiltonian 
is given by
\begin{eqnarray}
\label{Tube}
\hat {\cal H}
= \sum_{n=1}^N\sum_j[J\vec S_{n,j}\cdot\vec S_{n,j+1}
+J_\perp\vec S_{n,j}\cdot \vec S_{n+1,j}
-H S_{n,j}^z],
\end{eqnarray}
where $\vec S_{n,j}$ is the integer-spin operator on site $j$ in the
$n$th chain, $J(>0)$ and $J_\perp$ are the exchange couplings along 
the chain and rung directions 
($\vec S_{N+1,j}=\vec S_{1,j}$), respectively, and 
$H$ is the external field.
When the leg number $N$ is odd and the
rung coupling is antiferromagnetic (AF), i.e., $J_\perp>0$, 
the system exhibits frustration along the rungs. 
Our interest is in such frustrated cases.

For $N$-leg integer-spin systems (ladders and tubes), we can develop an
extended S\'en\'echal's nonlinear sigma model (NLSM) method~\cite{Sene}, 
which first maps each chain in the $N$-leg system to a
NLSM, and then treats the rung coupling terms perturbatively~\cite{M_S}. 
Although this method would be
efficient, especially in the weak rung-coupling regime, 
we believe that it is valid 
even in an intermediate coupling regime ($|J_\perp|\sim J$), 
at least qualitatively. 
The method shows that the rung coupling induces 
hybridization among the massive magnon bands in neighboring chains, 
and that magnon-band splitting occurs. 
All the resultant bands still have a finite gap.
In tubes, each band has a wave number $k$ for the rung direction.
The NLSM method also shows that the lowest magnon excitations 
in frustrated tubes possess a two-fold degeneracy 
in addition to the spin-1 magnon triplet. 
This extra degeneracy is guaranteed by the 
$\pi$-rotation symmetry with respect to the center 
axis of the cross section of the tube 
(see Fig. 3 in the first paper of Ref.~\cite{M_S}). 
Note that the symmetry is absent in ladders.
Here, recall that (as mentioned above) a sufficiently
strong field $H$ yields the condensation of $S^z=1$ magnons.
Therefore, one can immediately expect that a two-component TLL state 
appears as a result of the condensation of the lowest 
doubly degenerate magnons in the frustrated tubes. 
However, such a critical phase may actually be
broken down by the interactions between the two massless modes in the
condensed state.

To discuss the interaction effects, we use some field
theory approaches in addition to the NLSM. Applying a Ginzburg-Landau 
analysis for the field-induced critical phase in 
the spin-1 AF chain~\cite{Aff} to our NLSM theory in frustrated tubes, 
we obtain the following low-energy Lagrangian for
the two-component TLL expected above: 
${\cal L}_{\rm TLL} = \sum_{q=\pm p}
\frac{K}{2v}[(\partial_t\theta_q)^2-v^2(\partial_x\theta_q)^2]$.
The indices $q=\pm p$ represent the rung-direction 
wave number of each massless mode in 
the two-component TLL, $K$ is the TLL parameter~\cite{note1}, 
and $v$ is the velocity of the massless modes. 
The bosonic field $\theta_q(x)$ is associated with the spin operator 
$\vec S_{n,j}$ as follows: 
$\tilde S_{0,{\rm uni}}^z \sim \sum_{q=\pm p} 
[a \partial_x\phi_q/\sqrt{\pi}+M   
+ C_u \cos(\sqrt{4\pi}\phi_q+2\pi Mx/a)]$ 
and $\tilde S_{q,{\rm stag}}^+ \sim C_{1s} e^{i\sqrt{\pi}\theta_q}
[1  + C_{2s} \sin(\sqrt{4\pi}\phi_q+2\pi Mx/a)]$. 
Here $\tilde S_{k,j}^\alpha =\tilde S_{k,{\rm uni}}^\alpha 
+(-1)^j\tilde S_{k,{\rm stag}}^\alpha$ is 
the ``Fourier transformation'' of the original spin $S_{n,j}^\alpha$,
the field $\phi_q$ is the dual of $\theta_q$, 
$a$ is the lattice spacing ($x=j\times a$), 
$M$ is proportional to the magnetization per site, and
$C_{u,1s,2s}$ are the nonuniversal constants. 
The cos (sin) term on the right-hand side in 
$\tilde S_{0(q), {\rm uni}({\rm stag})}^{z(+)}$ is
expected from the study of the two-leg spin-1/2 ladder~\cite{FuZh}.
Note that we have not yet taken into account the interactions between 
the two TLLs, i.e., the fields $(\theta_p,\phi_p)$ and 
$(\theta_{-p},\phi_{-p})$.
In the bosonization plus renormalization group
picture~\cite{Gogo}, 
the relevant interaction terms are always 
represented by a product of some vertex operators, e.g., 
$e^{\pm i\sqrt{4\pi}\phi_q}$ and 
$e^{\pm i\sqrt{\pi}\theta_q}$~\cite{note2}. 
Relying on the symmetries~\cite{OYA} in frustrated tubes, 
we can restrict possible interaction forms in the low-energy theory 
as follows. From the form of $\tilde S_{q, {\rm stag}}^+$, 
the $U(1)$ rotation around the $z$ axis corresponds to the shift
$\theta_q\to \theta_q+{\rm const}$. This means that the
low-energy theory must not have any vertex operators, including
$e^{iC_1\theta_q}$ or $e^{iC_2(\theta_p+\theta_{-p})}$ ($C_{1,2}$ is a 
constant). Moreover, $\tilde S_{k,j}^\alpha$ tells us that 
the one-site translation along the chain corresponds
to $\phi_q\to \phi_q+\sqrt{\pi}M$ and 
$\theta_q\to\theta_q+\sqrt{\pi}$. 
We hence see the absence of all the vertex terms with 
$e^{iC_1\phi_q}$ or $e^{iC_2(\phi_p+\phi_{-p})}$, 
except for the case where $M$ is a specific commensurate value~\cite{OYA}. 
The remaining possible terms
are only $e^{iC_1(\theta_p-\theta_{-p})}$ and 
$e^{iC_1(\phi_p-\phi_{-p})}$. In addition to the above two symmetries,
the tube possess other ones: a translational symmetry along the rung
direction and a $\pi$-rotational one.
Employing these two, 
we can finally predict that these remaining terms are also 
prohibited in the low-energy theory. 
Thus, we conclude that a $c=1+1$ state emerges as the
lowest doubly degenerate magnons are condensed in the frustrated tubes.
The above argument also predicts the existence of a $c=1$ state in
non-frustrated $N$-leg systems.

This symmetry argument makes the
$c=1+1$ state strongly stabilized. Therefore, a natural question 
occurs: does the $c=1+1$ phase continue until the saturation of the
magnetization? To answer this, we still apply the above field theory
strategy to the case with a stronger field $H$ 
(although the original NLSM method is reliable in the zero-field case). 
In such a case, higher-energy magnons are condensed as well as the
lowest-energy ones. When the condensation of the second lowest magnons 
with wave number $q'$ takes place, 
new bosonic fields $\theta_{q'}$ and $\phi_{q'}$ must appear similar
to the appearance of $\theta_q$ and $\phi_q$. Here, note that 
in the three-leg case, the second lowest magnon has no degeneracy, namely
$q'$ take a single value $p'$, while in the cases with a larger odd leg number, 
the second magnons are doubly degenerate, i.e., $q'=\pm p'$.
Vertex terms including
only $\theta_{q'}$ and $\phi_{q'}$ are not allowed in the low-energy
theory, just like the case of the $c=1+1$ state.  
These new fields, however, provide the possibility of the presence of 
relevant terms consisting of $\theta_{q}$, $\phi_{q}$, 
$\theta_{q'}$ and $\phi_{q'}$. In fact, we can find that in the three-leg 
[larger-leg] case,
$\sum_{q=\pm p}\cos(\sqrt{4\pi}(\theta_q-\theta_{p'}))$
[$\sum_{q=\pm p,q'=\pm p'}\cos(\sqrt{4\pi}(\theta_q-\theta_{q'}))$]
is allowed by all symmetries and is expected to be relevant. 
Furthermore, it is shown that if these terms are present, only
the center-of-mass field $\Phi_c=\phi_{p}+\phi_{-p}+\phi_{p'}(+\phi_{-p'})$
still remains massless.
Therefore, we can predict that the emergence of these terms 
causes a quantum phase transition from the $c=1+1$ phase to a new $c=1$ one. 
A magnetization cusp would be observed at the transition point, 
because the magnetic susceptibility is usually proportional to the
number of massless modes in 1D spin systems.

\begin{figure}
\label{GS_Mag}
\includegraphics[height=.122\textheight]{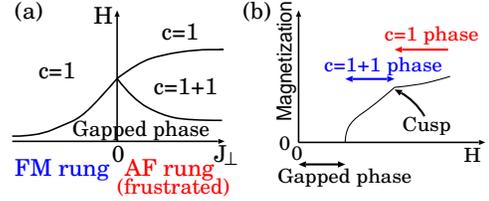}
\caption{Expected ground-state phase diagram (a) and magnetization curve 
(b) in the frustrated tube~(\ref{Tube}).}
\end{figure}

Summarizing the analysis of 
the frustrated integer-spin tube [Eq.~(\ref{Tube})], we can construct 
the ground-state phase diagram and the magnetization curve 
as in Fig.~\ref{GS_Mag}. 
Here, we should notice that there are a few subtle aspects
in the process leading to the $c=1+1$ phase and the
transition~\cite{M_S} (we omitted them in this paper). 
They might demand some modifications of our predictions.
Like the above case near the lower critical field, 
the arguments based on the symmetries would also be
powerful near the upper critical field. 
We will discuss such a high-field case elsewhere. 

The author would like to thank Masaki Oshikawa and Ian Affleck for 
valuable discussions.









\bibliographystyle{aipproc}   

\bibliography{sample}

\begin{thebibliography}{9}
\bibitem{note1}In the present notation~\cite{Gogo}, the scaling
	dimension of the vertex operator 
	$e^{\pm iA\theta_q}$ [$e^{\pm iA\phi_q}$] is $A^2/(4\pi K)$
	[$A^2K/(4\pi)$] ($A$ is a constant). The field $\phi_q$ is the
	dual field of $\theta_q$.
\bibitem{note2}Besides vertex operators, derivative terms such as
	$\partial_x\theta_q$ and $\partial_x\phi_q$ can be
	present. However, the first-order derivative terms can be
	absorbed by the Gaussian part ${\cal L}_{\rm TLL}$, and
	higher-order ones are at most marginal~\cite{Gogo}.
\end{thebibliography}

\IfFileExists{\jobname.bbl}{}
 {\typeout{}
  \typeout{******************************************}
  \typeout{** Please run "bibtex \jobname" to optain}
  \typeout{** the bibliography and then re-run LaTeX}
  \typeout{** twice to fix the references!}
  \typeout{******************************************}
  \typeout{}
 }

\end{document}